# Tunnel-Barrier-Engineered Ultrafast Demagnetization and Spin Transport in Graphene-Based Heterostructures

Suchetana Mukhopadhyay[1,2,†], David Muradas-Belinchón[3,†], Francesco Foggetti[3], Peter M. Oppeneer[3], M. Venkata Kamalakar[3,*], Anjan Barman[1,*]

[1]Department of Condensed Matter and Materials Physics, S. N. Bose National Center for Basic Sciences, Block JD, Sector III, Salt Lake, Kolkata 700106, India

[2]Department of Physical Sciences, Indian Institute of Science Education and Research Kolkata, Mohanpur, West Bengal 741252, India

[3]Department of Physics and Astronomy, Uppsala University, Box 516, Uppsala SE-751 20, Sweden

*Email: abarman@bose.res.in, venkata.mutta@physics.uu.se

[†]These authors have made equal contributions.

## ABSTRACT

Heterostructures combining graphene with $3d$ transition metal ferromagnets (FMs) enable various spin-based phenomena at ultrafast timescales. However, challenges such as the interfacial impedance mismatch, FM deposition-induced defect generation, and interface modification by interfacial coupling or hybridization can impede their functionalization for spin-orbitronics. In this work, we utilize insulating $TiO_x$ barrier layers (BLs) to modify the interfacial spin conductance structurally, disentangle spin pumping and magnetic proximity effects (MPE), and establish external control over ultrafast magnetization dynamics in single-layer graphene/$TiO_x$/Co systems. All-optical time-resolved magneto-optical Kerr effect measurements of femtosecond to nanosecond spin dynamics reveal systematic tunability of ultrafast magnetic parameters via barrier engineering. The thickness-dependent damping modulation in Co indicates strong spin pumping, with interfacial spin transparency close to half its physical limit in the presence of an ultrathin BL, where MPE is eliminated. Our results show that appropriately chosen ultrathin BLs can prevent interfacial alterations from ferromagnetic metals, facilitating efficient spin detection in graphene and enhancing control over spin angular momentum dissipation in graphene/FM interfaces.

## INTRODUCTION

The quest for realizing ultrahigh-speed spintronics technologies has followed a two-pronged approach: uncovering ultrafast magnetization manipulation mechanisms and engineering novel material systems for this purpose. Two decades since its experimental isolation, graphene, a two-dimensional (2D) atomic crystal of carbon[1], continues to stimulate both academic and industrial interest owing to its unprecedented multifunctionality, outstanding mechanical and electronic properties, and stability under ambient conditions[2, 3]. The weak intrinsic spin-orbit coupling (SOC) and negligible hyperfine interaction in graphene make it an excellent spin channel material showing micrometer-order spin diffusion length and spin lifetimes ~ns, and an ideal spin filter[4, 5]. These features, coupled with superior room-temperature electron mobility, gate-tunable carrier concentration[2, 6], and unmatched flexibility[7] make graphene an extremely promising candidate for advanced spintronics applications. Considerable strides have been made in the deposition and transfer of large-area graphene over the years[8], facilitating the integration of graphene in layered thin film heterostructures and spintronic devices[5, 7, 9]. Various reports have highlighted the combination of graphene with ferromagnetic thin films in spin valve or bilayer structures[6, 10] for observing various spin-orbit effects. The interfacial phenomena occurring in graphene/ferromagnet (FM) heterostructures generate additional control knobs for tuning the intrinsic electronic structure of graphene, leading to significantly altered magnetic and transport properties of the composite system[11].

Disentangling various charge and spin transfer processes occurring at the interface of graphene with a 3$d$ transition metal FM, such as Co, can be nontrivial[10, 11]. Interaction of the Co-$d_z^2$ orbitals with the out-of-plane $p\pi$ bands of the constituent $sp^2$ carbon of graphene results in strong interfacial $pd$ hybridization and charge transfer[12]. On the other hand, spin angular momentum transfer can be achieved by injecting spin-polarized carriers across the interface. However, the efficiency of spin injection occurring via carrier transport is fundamentally limited by interfacial impedance mismatch[13, 14]. Dynamical spin pumping induced by magnetization precession in an FM does not involve a net charge flow and hence bypasses the impedance mismatch issue[15, 16], requiring only the interfacing nonmagnet to be a good spin sink[15, 17, 18]. Despite having low intrinsic SOC strength in its pristine form, nearly 20-fold SOC enhancement of extrinsic origin can occur in graphene deposited on substrates due to the unavoidable rippling of the graphene sheet[19] and introduction of impurities or adatoms[20]. Consequently, graphene interfaced with an FM thin film can act as an efficient sink for the dynamically generated pure spin current from the precessing FM magnetization[13, 21]. However, although high spin conductance values have been reported, interfacial hybridization-induced magnetic proximity effects (MPE) typically occur concomitantly with the spin transport in graphene/FM heterostructures[22, 23], which compromises interface integrity and consequently the utility of graphene as a spin detector.

Ultrafast optical excitations can effectively induce magnetization precession, and signatures of dynamical spin pumping can manifest at ultrashort timescales down to a few picoseconds[24, 25]. The optically induced subpicosecond magnetic phenomenon known as ultrafast demagnetization has attracted immense research interest for being the fastest known mechanism for manipulating magnetic order easily realized in room-temperature experiments. While the introduction of tunnel barriers at graphene/FM interfaces is well known to alleviate the impedance mismatch and enhance electrical spin-injection efficiency[26], their effect on the precession-induced spin pumping and ultrafast demagnetization in graphene/FM systems remains to be studied. In the present work, we used TiO$_x$ barrier layers (BLs) to tune the interfacial coupling and spin transport in a strongly coupled graphene/Co bilayer system. Varying the barrier thickness allows significant control over the subpicosecond ultrafast demagnetization and nanosecond precessional dynamics by modulating the MPE and spin transparency at the interface. Our graphene/TiO$_x$/Co heterostructures, grown by a combination of chemical vapor deposition (CVD) and electron beam evaporation (EBE), exhibit high quality with no evidence of structural compromise due to the deposition of the oxide BL or the FM. By performing all-optical magnetization dynamics measurements on these engineered samples, we uncover the dependence of magnetic damping and ultrafast demagnetization time on the BL thickness. We find that the damping enhancement in directly contacted graphene/Co decreases as the BL thickness is increased, allowing us to identify a length scale across which pure spin current transport persists in the system. This behavior is corroborated by superdiffusive spin transport calculations that account for the transfer of spin current across the BL. The Co thickness-dependent modulation of magnetic damping bears the signatures of a strong dynamical spin pumping occurring at the graphene/Co interface, with a high spin-mixing conductance even in the presence of an ultrathin TiO$_x$ barrier. Our work unambiguously demonstrates efficient spin pumping across a tunneling barrier in a graphene/FM system without interfacial modification of graphene. The dependence of the magnetic parameters on the BL thickness reveals an avenue for controlling spin angular momentum transfer across 2D material/FM thin film heterostructures, offering a promising new route to engineer their functionality and applicability, advancing ultrafast and interface-protected spintronic device architectures.

**RESULTS AND DISCUSSION**
**Sample growth and characterization**
$TiO_x$ BLs of various thicknesses ($\delta$ = 0.8, 1.6, 2.4, & 5 nm) followed by a FM Co layer were deposited by EBE on large-area CVD-grown graphene on $Si/SiO_2$ substrate. CVD is well known as the most reliable and industry-compatible technique for large-area graphene deposition [8, 27]. Subsequently, the films were capped with 4-nm-thick Ti which forms a native oxide layer, preventing oxidation of the Co surface. Deposition of the oxide and FM layer can introduce defects in the underlying graphene and potentially compromise its structural quality. The presence of defects in graphene plays an important role in controlling spin transport at graphene/FM interfaces [28] in addition to modifying the Fermi energy and band structure[20, 29]. Consequently, a functional dependence of the defect density on the BL thickness $\delta$ would present a confounding factor in our interpretation of results, making it crucial to rule out this possibility before proceeding with our analysis. To this end, micro-Raman spectroscopy measurements ($\lambda$ = 532 nm, microscope objective: 100×, grating: 600 lines/mm, N.A. = 0.9 and spot diameter ≈ 0.7 μm) were carried out on the graphene/$TiO_x$/Co heterostructures to ascertain the quality of the underlying graphene and quantify the defect density. Raman spectroscopy is one of the most powerful non-destructive techniques for characterization of carbon-based materials, such as graphene, and is particularly well adapted for detecting structural changes in carbon nanomaterials[30].

The characteristic features in the Raman spectra of graphene are the first-order $G$ band which arises due to doubly-degenerate optical phonon modes at the Brillouin zone center, and the $2D$ band arising from a second-order scattering involving two optical phonons at the K point[31, 32]. Raman shifts obtained from the graphene/$TiO_x(\delta)$/Co($d$ = 5 nm) films are presented in Figure 1a, where $G$ band is observed at a shift of 1585 cm$^{-1}$ and $2D$ band at 2680 cm$^{-1}$. The inset of the figure shows the perfect single Lorentzian lineshape of the $2D$ band for graphene/Co, which indicates that the underlying graphene is single-layered (hereafter referred to as SLG: single-layer graphene) [32]. In the presence of disorder in graphene, a third so-called "$D$" band emerges from a second-order scattering involving a phonon and a defect center at around half the frequency of the $2D$ band. We find that the $D$ band intensity remains substantially low for all our samples indicating that EBE deposition of the $TiO_x$ BL and Co metal does not compromise the quality of the underlying graphene. This is in contrast to sputtered FM deposition, which can induce substantial disorder into the graphene lattice [33]. The defect density in the graphene can be further quantified using the relative intensity ratio of the $D$ band and $G$ band peaks ($I_D/I_G$). The average crystallite size ($L_a$) in graphene, which is a measure of the average distance between neighboring defects or the defect density, is calculated as a function of $I_D/I_G$ ratio, as [34]

$$L_a = (2.4 \times 10^{-10})\lambda^4 \left(\frac{I_D}{I_G}\right)^{-1} \qquad (1)$$

As shown in Figure 1c, the $I_D/I_G$ ratio or the defect density $L_a$ does not show any systematic dependence on the BL thickness, further verifying that the deposition of the $TiO_x$ and Co layers does not compromise the structural integrity of the SLG or increase the defect density. Further, as seen from Figure S2 of the Supplementary Information, variation of Co layer thickness does not alter the Raman characteristics. Hence, any possible role played by impurity or defect scattering in the multilayer samples is minor. Moreover, the BL-thickness and Co-thickness-dependent analyses can be expected to be unaffected by variations in defect density. Having established the good structural quality of the underlying graphene in all the SLG/TiO$_x$/Co samples, the surface topography of the samples was studied using atomic force microscopy (AFM) measurements. The average topographical roughness of the films at very low thicknesses exhibits the signature of interfacial roughness. AFM images of SLG/TiO$_x$/Co films are shown in Figure 1d with the average topographical roughness $S_a$ values quoted in Table 1[35]. The $S_a$ value is found to be relatively low in the sub-nm range for all samples, regardless of TiO$_x$ thickness. The negligible variation in roughness values measured at different regions of the same sample indicates uniform film growth. Notably, the $S_a$ value for the Co sample grown directly on Si/SiO$_2$ does not differ significantly from those deposited on SLG/TiO$_x$. In order to investigate the wetting of TiO$_x$ on SLG prior to Co deposition, AFM topography measurements of SLG before and after deposition of TiO$_x$ barrier layers with various nominal thicknesses were also carried out as shown in Figure S4 of the Supplementary Information, in addition to that of the complete stack after deposition of Co and capping layer.

Table 1: Average topographical roughness $S_a$ extracted from AFM measurements.

| $\delta$ (nm) | No SLG | 0 | 0.8 | 1.6 | 2.4 | 5 |
|---|---|---|---|---|---|---|
| $S_a$ (nm) | 0.18 | 0.22 | 0.22 | 0.22 | 0.26 | 0.24 |

**Magnetization dynamics**

We proceeded to carry out time-resolved magneto-optical Kerr effect (TRMOKE) measurements of the femtosecond laser-induced ultrafast magnetization dynamics in the SLG/TiO$_x$/Co samples. A schematic of the two-color pump-probe measurement geometry using a 400 nm pump and 800 nm probe beam is shown in Figure 2a. Prior to laser excitation, an in-plane bias magnetic field of 1.9 kOe is applied at a slight tilt of 10°-15° to saturate the magnetization of the samples. Exciting a FM material with a femtosecond laser pulse results in a rapid drop of the magnetization at sub-picosecond timescales, the phenomenon known as ultrafast demagnetization[25]. Following this ultrafast quenching, the magnetization recovers over two characteristic timescales, a fast recovery process typically occurring over a few picoseconds, and a slow recovery that persists well into the nanosecond time regime coinciding with the damping of the precessional oscillations. The coherent magnetization precession is triggered by a laser heating-induced ultrafast anisotropy change[36] and damps out on a characteristic, material-dependent, timescale. A typical time-resolved Kerr rotation trace is shown in Figure 2b for the SLG/Co sample measured with a pump fluence of ≈10 mJ cm$^{-2}$. The precessional Kerr rotation signals measured from the SLG/TiO$_x$($\delta$)/Co(5 nm) samples under an applied bias magnetic field of 1.9 kOe are shown in Figure 2c. The background-subtracted precessional data are fitted with a damped sinusoidal function to extract the relaxation time $\tau$ of the oscillations. A fast Fourier transform (FFT) is performed on the precessional data to obtain the central frequency of the fundamental mode. The bias magnetic field dispersions of the central frequency values are fitted to the Kittel dispersion for FM systems to extract the effective magnetization $M_{eff}$ of the probed volume [37]

$$f = \frac{\gamma_0}{2\pi}\sqrt{H(H + 4\pi M_{eff})} \qquad (2)$$

where $H$ is the bias magnetic field and $\gamma_0$ is the gyromagnetic ratio. As compared with the result for the bare Co sample grown on Si/SiO$_2$, there is a noticeable redshift of the fundamental mode frequency in the presence of SLG underlayer, which is no longer observed upon the introduction of even an ultrathin 0.8-nm-thick TiO$_x$ BL. This redshift results in a lower $M_{eff}$ value as extracted from the Kittel dispersion relation for SLG/Co than that of the SLG/TiO$_x$($\delta\neq0$)/Co samples as shown in Figure 2f. We thus conclude that the mechanism leading to the significant reduction of Co moment in presence of SLG underlayer has been effectively isolated by the BL.

Figure 3a depicts the ultrafast demagnetization measured by TRMOKE for the SLG/TiO$_x$($\delta$)/Co(5 nm) samples. The characteristics of the demagnetization provides important clues for identifying the spin angular momentum relaxation channels present in the system[38]. Before quantifying the demagnetization timescales, a few critical observations can be made directly from the traces. Firstly, the demagnetization appears to proceed much faster in the presence of SLG. Secondly, the demagnetization time changes with the thickness $\delta$ of the TiO$_x$ BL. Finally, the degree of modulation of demagnetization time with $\delta$ is considerably more pronounced in these samples when compared to the SLG/TiO$_x$($\delta$)/Co(20 nm) samples discussed in Section I of the Supplementary Information. These factors indicate that the dependence of the demagnetization rate on $\delta$ is connected to a mechanism that is more prominent at lower Co thickness. The nonequilibrium situation prevailing immediately after the laser excitation is difficult to encapsulate in analytical treatments and a phenomenological approach is commonly preferred to extract quantitative information from demagnetization profiles. In the latter, the demagnetization is qualitatively explained by a three-temperature model which describes energy pumping into the electronic system and its subsequent redistribution among the spin and the lattice systems and simulates the ultrafast demagnetization as arising from a transient temperature rise of the spin subsystem. From this model, the following phenomenological expression for the transient magnetization at picosecond timescales can be derived [39]:

$$-\frac{\Delta M_z}{M_z} = \left[\left\{\frac{A_1}{\left(1+\frac{t}{\tau_0}\right)^{\frac{1}{2}}} - \frac{(A_2\tau_R - A_1\tau_M)e^{-\frac{t}{\tau_M}}}{\tau_R - \tau_M} - \frac{\tau_R(A_1 - A_2)e^{-\frac{t}{\tau_R}}}{\tau_R - \tau_M}\right\}\Theta(t) + A_3\delta(t)\right] * \Gamma(t) \qquad (3)$$

where $\Theta(t)$ is the Heaviside step function, $\delta(t)$ is the Dirac delta function, and $\Gamma(t)$ represents the Gaussian laser pulse. The amplitude $A_1$ represents the normalized magnetization amplitude at the end of the fast remagnetization process signaling re-equilibration between the electron, lattice and spin systems. $A_2$ is proportional to the initial rise of electron temperature and hence the maximum magnetization quenching, while $A_3$ represents the magnitude of state-filling effects at the onset of demagnetization. $\tau_0$, $\tau_M$, and $\tau_R$ are time constants representing the heat diffusion timescale, the ultrafast demagnetization time, and the fast magnetization recovery time, respectively. The experimentally measured demagnetization traces are fitted to Equation 3 to extract the characteristic demagnetization timescale $\tau_M$ as well as $\tau_R$. Figure 3b shows the dependence of $\tau_M$ and $\tau_R$ on the BL thickness $\delta$ for SLG/TiO$_x$($\delta$)/Co(5 nm) samples. The remagnetization proceeds faster for SLG/Co than for SLG/TiO$_x$/Co, which is most likely due to the high in-plane thermal conductivity of graphene. The weak dependence of $\tau_R$ on $\delta$ for the thin Co samples is due to the closer proximity to the substrate for thinner BLs. Likewise, the smaller $\tau_R$ values for the 5 nm Co samples compared to the 20 nm ones as seen from Figure S1c of the Supplementary Information can also be attributed to the proximity to the substrate which has a higher thermal conductivity than the Co. The extracted $\tau_M$ values confirm the two observations discussed above, both of which indicate the presence of an additional spin relaxation channel in SLG/Co. Firstly, the significant reduction of $\tau_M$ in the presence of SLG, which serves as the first indicator that the SLG underlayer acts as a spin sink, providing a channel for spin angular momentum transfer and enhancing the demagnetization rate[40, 41]. Secondly, for 5-nm-thick Co, a clear decrement of $\tau_M$ with $\delta$ is observed before reaching a saturation. Finally, the modulation of $\tau_M$ with $\delta$ is more prominent in the samples having $d$ = 5 nm versus $d$ = 20 nm. As shown in Figure 3b, for Co directly laid on SLG, a strong acceleration of the ultrafast demagnetization is observed from the reference Co sample. As the BL is introduced, the relative decrease in $\tau_M$ is diminished which further decreases with increasing $\delta$, approaching the value of $\tau_M$ in the reference Co sample for $\delta$ = 5 nm. This indicates that the spin angular momentum transfer is effectively attenuated by increasing the TiO$_x$ thickness. As shown in Figure S1 of the Supplementary Information, a similar trend of $\tau_M$ with $\delta$ is observed for 20 nm Co, although the demagnetization is comparatively slower. Thus, a large degree of tunability of demagnetization time has been achieved in our SLG/Co samples, which turns out to be easily controllable by varying the thickness of a TiO$_x$ BL.

Figure 3c shows the modulation of the Gilbert damping parameter $\alpha_{eff}$ as a function of the BL thickness $\delta$, calculated using the relation $\alpha_{eff} = 1/[\gamma_0\tau(H+2\pi M_{eff})]$. The large enhancement of damping in the presence of SLG underlayer can arise due to the injection of the pure spin current generated by the magnetization precession in Co and its dissipation in the SLG. Interestingly, a considerable damping enhancement is also observed for the sample with the 0.8-nm-thick BL, indicating spin current dissipation still takes place in the presence of the ultrathin insulating spacer. The dependence of $\alpha_{eff}$ on $\delta$ can be fit well to an exponential decay, revealing a length-scale $t_{BL}$ = 0.83 nm associated with the persistence of spin relaxation across the TiO$_x$ BL. We note that this is not immediately anticipated, as enhancement of spin pumping-induced voltage signals by applying insulating tunnel barriers has also been reported [42]. Further, the degree of damping modulation achieved by varying $\delta$ is considerably larger for these samples than for the SLG/TiO$_x$($\delta$)/Co(20 nm) set discussed in the Supplementary Information. The fact that the variation of both $\tau_M$ and $\alpha_{eff}$ with the BL thickness is considerably less pronounced for samples with $d$ = 20 nm as opposed to $d$ = 5 nm, serves as a prime indicator for the interfacial origin of the modulation of ultrafast magnetic properties as a function of $\delta$. This less pronounced $\delta$-dependence for the samples having thicker Co cannot be due to MPE, as it is independent of Co thickness and is already eliminated by the thinnest BL. This leads us to infer that it is instead connected to a distinct, thickness-dependent process that is more prominent at lower FM thickness. Spin-pumping-induced damping enhancement follows a characteristic inverse dependence on the FM layer thickness[17]. To investigate whether such a functional relationship exists for our samples, we additionally fabricated two sets of samples, SLG/Co($d$) and SLG/TiO$_x$($\delta$ = 0.8 nm)/Co($d$), where $d$ = 2, 3, 4, 5 & 10 nm. The observed FM thickness-dependent damping shown in Figure 4a can be well described by the functional relationship.

$$\alpha_{eff} = \alpha_0 + \frac{g\mu_B}{4\pi M_s d}G_{eff} + \frac{\beta_{TMS}}{d^2} \qquad (4)$$

as seen from the fitted curves in the figure. Here, $G_{eff}$ is the effective spin-mixing conductance quantifying the efficiency of pure spin current transmission driven by the dynamical spin pumping, while $\beta_{TMS}$ quantifies the interfacial losses incurred due to the dephasing of the uniform ferromagnetic resonance mode in the presence of interfacial magnetic roughness via the two-magnon scattering (TMS) mechanism[43]. Thus, the sharp near-linear $d^{-1}$ dependence of damping is indicative of strong spin pumping occurring in the presence of SLG, while TMS contributes a characteristic parabolic scaling with $d^{-1}$. The TMS strength is quantified by the coefficient $\beta_{TMS}$, which is reduced from $(3.8 \pm 0.4) \times 10^{-17}$ cm$^2$ for SLG/Co to $(1.7 \pm 0.04) \times 10^{-17}$ cm$^2$ in the presence of TiO$_x$. Notably, although the $G_{eff}$ value is reduced from $(5.1 \pm 0.02) \times 10^{15}$ cm$^{-2}$ in SLG/Co to $(3.2 \pm 0.01) \times 10^{15}$ cm$^{-2}$ with the introduction of the 0.8-nm-thick TiO$_x$ BL, it still remains sizeable, indicating that pure spin current transport persists via the ultrathin BL.

There is a pressing need for comprehensive theoretical treatments of spin pumping in 2D material/FM heterostructures to interpret experimental results better. Care must be taken while estimating the intrinsic spin-mixing conductance from the experimentally determined $G_{eff}$ values[17, 18] as the spin diffusion length (SDL $\lambda_s$) considered in the model applies to transverse spin current transport occurring perpendicular to the interface. Consequently, using the in-plane micrometer order SDL of graphene for spin pumping would erroneously result in a severe overestimation of interfacial spin backflow. According to Ohshima et al.[44], the in-plane $\lambda_s$ of graphene (~ 2 μm) must be reduced by a factor of $3.75 \times 10^3$, derived from a relation connecting the in-plane and transverse conductivities in graphite[45]. This produces the estimated $\lambda_s = 0.53$ nm for vertical spin transport [44], indicating that SLG can indeed sink the pumped spin current from the FM, which can subsequently dissipate over micrometer-order length-scales in the graphene plane. Using the extracted $G_{eff}$ values and the calculated $\lambda_s = 0.53$ nm, the intrinsic spin-mixing conductance $G_{\uparrow\downarrow}$ of the SLG/Co and SLG/TiO$_x$/Co samples are extracted using the expression

$$G_{\uparrow\downarrow} = G_{eff} \left(1 - e^{\frac{-2t}{\lambda_s}}\right)^{-1} \qquad (5)$$

assuming quasi-ballistic spin transport through the thin SLG layer ($t < \lambda_s$)[46]. As discussed in the Supplementary Information, it is a more complicated affair to estimate the interfacial spin transparency[47] in SLG/Co (or 2D material/FM heterostructures in general), by making use of the existing theoretical models employed in the literature. Taking $t = 0.335$ nm[48] as the thickness of SLG, we obtain $G_{\uparrow\downarrow} = (7.1 \pm 0.02) \times 10^{15}$ cm$^{-2}$ for SLG/Co and $G_{\uparrow\downarrow} = (4.5 \pm 0.01) \times 10^{15}$ cm$^{-2}$ for SLG/TiO$_x$(0.8 nm)/Co, indicating that a significant spin pumping effect takes place in the presence of the ultrathin BL. Finally, the ultrafast demagnetization and Gilbert damping correlation are probed as a function of Co layer thickness $d$. The demagnetization time $\tau_M$ increases with $d$, resulting in an inversely proportional relationship between $\tau_M$ and $\alpha_{eff}$ as shown in Figure 4b, further confirming the presence of additional spin relaxation channels arising from the spin pumping effect[38, 41].

The above results can be understood in the context of existing literature on various interfacial phenomena at play in graphene/metal heterostructures. The nature of the interactions between the out-of-plane $p\pi$ states of graphene and the states of the interfacing metal mediates the change in the electronic states of graphene near the Fermi level[49]. The strength of the bonding between graphene and metal can be classified as weak or strong depending on the extent of modification to the pristine electronic structure of graphene. In weakly interacting systems where the graphene-metal distance is comparable to the interlayer distance in graphite[50], the linear energy dispersion and Dirac cone of graphene are essentially unperturbed[51], and the graphene π bands show no signs of hybridization. On the other hand, in strongly interacting systems, the Dirac cone is modified and the π bands are shifted to higher binding energies[50, 52]. Directly contacted Co (001) or Ni (111) on graphene are prototypical systems for the latter case, in which strong hybridization of the graphene π bands with the Co $d_z^2$ or Ni $d_z^2$ orbitals leads to the appearance of a hybridization gap in the graphene band structure and a notable π band shift[53], though the preservation of the Dirac cone is still debated [53-55]. Proximity coupling to FM layers via a direct interface can induce a magnetic character in the electronic bands of graphene in strongly interacting systems[23, 56]. The hybridization between graphene π and spin-polarized 3d bands of the FM leads to induced spin polarization of the π states in graphene, resulting in an effective magnetic moment of the carbon with a lowering of the magnetic moment of the FM [56, 57]. Although there have been some debates on the origin of this MPE [58], the appearance of magnetic moments in carbon π electrons in a Ni/graphene system has been unambiguously demonstrated[54, 59]. Co has a three-times larger 3d magnetic moment than Ni, and can therefore enhance the spin polarization of the graphene π bands. Thus, the redshift in precessional frequency and lower $M_{eff}$ value observed in our SLG/Co samples can be attributed to the MPE arising from interfacial sp-d hybridization, which is eliminated by the insertion of $TiO_x$. Notably, while the BL effectively isolates the MPE, it does not affect the magnetic surface anisotropy appreciably, as seen from the minor variation of $M_{eff}$ with increasing BL thickness.

The presence of MPE arising from sp-d hybridization can potentially enhance the magnetic damping[60]. Meanwhile, the extrinsic TMS contribution to the damping modulation for 2 nm Co is around $(3.8 \pm 0.4) \times 10^{-17}$ cm$^2$ for SLG/Co which is decreased to $(1.7 \pm 0.04) \times 10^{-17}$ cm$^2$ in the presence of the $TiO_x$ BL. However, since 0.8-nm-thick TiOx already eliminates the MPE, any additional damping arising from MPE cannot explain the variation shown in Figure 3c as a function of BL thickness. On the other hand, since the TMS contribution to the damping modulation is minor, TMS alone cannot account for the difference in damping between SLG/Co and SLG/$TiO_x$/Co. While thin insulating BLs can improve spin injection efficiency via tunneling spin-polarized carrier transport, a transparent interface is most conducive to the transport of pure spin current generated by dynamical spin pumping associated with the damping enhancement. The modulation of damping as a function of FM layer thickness provides the most important evidence for the origin of the damping enhancement in SLG/$TiO_x$/Co, proving that although the interface transparency is reduced, spin pumping persists across the 0.8-nm-thick BL. As the TiOx thickness increases, spin pumping becomes less effective, and spin injection is reduced, resulting in a decrease in the Gilbert damping. The Co thickness dependence of demagnetization time and the modulation of demagnetization rate by the BL thickness are observations we now attribute to the interfacial nature of the spin pumping phenomenon and the attenuation of pure spin transport by the $TiO_x$ BL. Figure 3 also reveals an inverse relationship between demagnetization time and Gilbert damping as a function of BL thickness, further confirming that the BL modulates spin relaxation. The sizable spin-mixing conductance and moderate interfacial transparency for SLG/$TiO_x$/Co indicate that the ultrathin tunnel barrier at the SLG/Co interface allows spin transport while eliminating interfacial hybridization, thereby allowing for the isolation of these two effects in practice. As the presence of interfacial coupling and MPE impedes the spin detection functionality of graphene in SLG/FM heterostructures, the elimination of this effect by introducing adequately thin $TiO_x$ BLs holds promise for spin-orbitronic applications.

**Superdiffusive spin transport simulations**

To support our results, we carried out numerical simulations of laser-induced demagnetization in a 5 nm Co layer, whose interface properties were varied according to the presence of different thicknesses of $TiO_x$ BLs. To do so, we used the superdiffusive spin transport equation[24, 61] (see Methods), given by:

$$\frac{\partial n_\sigma}{\partial t} + \frac{n_\sigma}{\tau_\sigma} = \left(-\frac{\partial}{\partial t}\hat{\varphi} + \hat{I}\right)\left(\hat{S}_\sigma^{eff}\right) \quad (6)$$

where $n_\sigma$ is the density of laser-excited electrons, $\tau_\sigma$ represents the lifetime of excited electrons, $\hat{I}$ is the identity operator, and $S^{eff}$ is the effective spin source that takes into account the Gaussian-like profile of excited electrons as well as the electrons generated in the material by scattering events. The effects of the $TiO_x$ barrier are taken into account by the action of the flux operator $\hat{\varphi}$, which contains information on the reflection properties of electrons inside the material and at the interfaces. Solving Equation (6) allows us to obtain the demagnetization profile for the Co layer we are considering, from which the demagnetization time $\tau_M$ is estimated. By modeling the transmission properties of the Co/$TiO_x$ interface, we can compute the reflection coefficients at the interface with the $TiO_x$ barrier for different thicknesses of the barrier and use them to evaluate the thickness-dependent demagnetization time in Co.

We model the transmission coefficient for a particle of energy $E$ through a one-dimensional barrier of energy $V$ and thickness $\delta$

$$T(E,\delta) = \frac{1}{1 + V^2 \sinh\alpha\delta^2/\bigl(4E(V-E)\bigr)}, \quad (7)$$

Here, we choose the energy barrier $V$ to be equal to the $TiO_2$ bandgap, reported as 3.7 eV in the literature[62], while $\alpha$ is extrapolated from the experimental results given in Figure 3c. This equation provides the thickness-dependent reflection coefficients needed to simulate the demagnetization dynamics in the Co/$TiO_x$ systems for different thicknesses of the $TiO_x$ BL; the computed results are shown in Figure 5.

Comparing Figure 5 with the experimental results given in Figure 3b, we observe a good qualitative agreement: the demagnetization time grows initially with the $TiO_x$ BL thickness and then saturates for thicknesses larger than 4 nm, exhibiting the same trend as the experiment. A quantitative difference is observed in the demagnetization time $\tau_M$ which is faster in the experiment than in the simulation. This might be because the superdiffusion model predicts laser-induced demagnetization but does not include the remagnetization dynamics. The simulated demagnetization time in Figure 5 is moreover computed by approximating the magnetization profile as a simple decaying exponential $M(t) \approx M(0)\exp(-t/\tau_M)$, whereas the demagnetization time $\tau_M$ reported in Figure 3b is extracted using the phenomenological expression in Equation (3), which takes the remagnetization into account.

The use of graphene/FM systems is fundamental to 2D spintronics, in particular for spin-polarized transport in graphene. Proximity effect and contact-induced spin relaxation are issues in graphene, and direct contacts often result in additional spin relaxation, presenting a roadblock to realizing high-efficiency spintronic applications such as graphene spin torque oscillators and high-frequency devices. Our work describes a model system for the field of 2D spintronics and magnetism, providing ultrafast optical and dynamical confirmation that a thin BL permits efficient spin pumping across a graphene/FM interface while decoupling MPE, which is essential for ultrafast spin logic, optical switching, and spin-based signal processing at GHz–THz frequencies. At the same time, our results provide a basis for investigations into other layered materials and van der Waals heterostructures (incorporating transition metal dichalcogenides, magnetic topological insulators, etc.) where interfacial spin transparency, damping, and MPE may be selectively tuned. It advances the experimental toolkit for picosecond-nanosecond spin dynamics by combining time-resolved Kerr-based magnetometry, damping measurements, and spin transport modeling, thus contributing a framework for benchmarking interfacial spin relaxation in emerging 2D systems.

**CONCLUSIONS**
We demonstrate efficient spin pumping into single-layer graphene (SLG) through Co, with systematic control of ultrafast magnetic dynamics enabled by ultrathin $TiO_x$ barrier layers (BLs). The $SLG/TiO_x/Co$ heterostructures, fabricated via electron beam evaporation (EBE), preserve graphene integrity, compared to sputtered approaches. Time-resolved magneto-optical Kerr effect (TRMOKE) measurements reveal that SLG enhances demagnetization rate ($1/\tau_M$) and Gilbert damping ($\alpha_{eff}$) in Co, and these effects remain significant even after introducing a 0.8 nm $TiO_x$ BL, which effectively suppresses the magnetic proximity effect (MPE) at the SLG/Co interface. The observed reciprocal, nonmonotonic trends in $\tau_M$ and $\alpha_{eff}$ with barrier thickness are supported by superdiffusive spin transport modeling. The Co thickness-dependent damping confirms the interfacial nature of spin pumping, with the extracted spin mixing conductance values confirming efficient spin transfer despite the insulating spacer. The ability to maintain spin pumping across an insulating layer, while independently suppressing proximity effects is experimentally nontrivial and has not been previously achieved in such systems. Thus, our approach establishes a direct method to decouple MPE from spin injection, which is significant for spintronic applications where temporal resolution and interface control are key. These findings enhance the understanding of spin transport in 2D material/FM systems and provide a functional strategy for designing scalable, ultrafast spintronic devices based on graphene.

**METHODS**
**I. Sample fabrication**
CVD-grown SLG on Cu foil substrate was commercially procured from Graphenea Inc. The SLG was transferred using a semi-dry transfer process to a previously patterned 4" Si wafer with a 285-nm-thick $SiO_2$ layer. The wafer was then coated with a protective resist and cut into 7.3 mm × 7.3 mm chips. After cleaning the chips to remove the protecting resist, $TiO_x$ barrier layer (BL) of desired thickness was grown by deposition of Ti by EBE with a base pressure of $3 \times 10^{-7}$ Torr followed by oxidation at ambient conditions. Co layer of 5 nm or 20 nm thickness was subsequently deposited in the same chamber and capped with 4 nm Ti. The Ti oxidizes to form the native oxide layer which protects the underlying metal from oxidation. Very low deposition rates of 0.1 nms$^{-1}$ for Ti and 0.15 nm s$^{-1}$ for Co were used to achieve uniform deposition. Two sets of films with varying Co layer thickness were also deposited: the first set directly on the cleaned SLG and a second after growth of 0.8 nm $TiO_x$ BL. The detailed structural characterization of these samples is described in the SI.

**II. Time-resolved all-optical measurements**
The laser-induced magnetization dynamics in the SLG/Co samples were measured using a custom-built TRMOKE setup. The two-color detection scheme using a noncollinear measurement geometry is illustrated in Figure 2a[63]. The samples were pumped with optical pulses having λ = 400 nm and pulsewidth ~ 35 fs generated by frequency doubling the s-polarized 800 nm fundamental output of an amplified femtosecond laser (LIBRA, Coherent, pulsewidth = 35 fs, and repetition rate = 1 kHz). The optical pump-probe detection scheme enables the construction of the complete time-domain picture of the ultrafast magnetization dynamics and offers excellent time resolution limited only by the laser pulsewidth. In the noncollinear measurement geometry the exciting pump beam is obliquely incident on the sample while the probe beam is incident perpendicular to the sample plane. The pump beam and the probe beam were made to exactly coincide on the sample surface by careful optical alignment. The change in the transient magnetization of the sample in response to optical pumping is manifested as a change in the magneto-optical Kerr rotation of the time-delayed probe. A remotely operated variable delay generator offers precise control of the pump-probe time delay. A mechanical chopper operating at 373 Hz is kept in the path of the pump beam to provide the reference signal for phase-sensitive lock-in detection of the Kerr rotation signal. Prior to laser exposure, a large magnetic field of 1.9 kOe was applied at a tilt of about 10°-15° from the sample plane to saturate its magnetization. All experiments were carried out under ambient temperature and pressure conditions.

III. Simulation details

For simulating the laser-induced demagnetization in SLG/TiO$_x$/Co, the superdiffusion transport theory[24, 61] was used, wherein the time evolution of the density of laser-excited electrons $n_\sigma$ is described by Equation (6), a quasi-1D differential equation involving energy-, position- and time-dependent quantities where $\sigma$ represents electron spins (up or down). The effective spin source term $\hat{S}_\sigma^{eff}$ takes into account the Gaussian-like profile of excited electrons as well as the electrons generated in the material by scattering events. To simulate the excitation from the 400 nm laser pulse, electrons are excited up to 3 eV above Fermi level, in 24 discrete, uniformly spaced energy levels. The lifetimes and velocities of the electrons are taken from reference[64]. We use a decaying exponential to describe the number of excited electrons $N(z)$ at the depth coordinate $z$ in the material as $N(z) = N_0 \exp(-z/\lambda)$, where $N_0$ is the number of excited electrons in the first nm of material, and $\lambda = 21.58$ nm is the laser penetration depth that is computed by using the Beer's Law $\lambda = L/(4\pi k)$, where $L$ is the laser wavelength and $k$ the extinction coefficient of Co taken from reference[65].

**ACKNOWLEDGMENTS**
AB gratefully acknowledges S. N. Bose National Centre for Basic Sciences (grant no. SNB/AB/12-13/96) and Nano Mission, Department of Science and Technology (DST), Govt. of India (grant no. DST/NM/TUE/QM-3/2019-1C-SNB) for funding. SM acknowledges DST, India for financial support from INSPIRE fellowship. This research has made use of the Technical Research Centre (TRC) Instrument facilities of S. N. Bose National Centre for Basic Sciences, established under the TRC project of DST, Govt. of India. MVK gratefully acknowledges funding from European Research Council (ERC) Project SPINNER (Grant No. 101002772), Stiftelsen Olle Engkvist Byggmästare (200–0602), Energimyndigheten (48698–1), FLAG-ERA Project MINERVA Swedish Research Council VR (2021-05932), Formas (2023-01607), and the Knut and Alice Wallenberg Foundation (Grant No. 2022.0079). PMO acknowledges support by the Swedish Research Council (VR) and the Knut and Alice Wallenberg Foundation (Grants No. 2022.0079 and 2023.0336). The spin-transport calculations were enabled by resources provided by the National Academic Infrastructure for Supercomputing in Sweden (NAISS) at NSC Linköping, partially funded by VR through Grant Agreement No. 2022-06725.

**FIGURES**

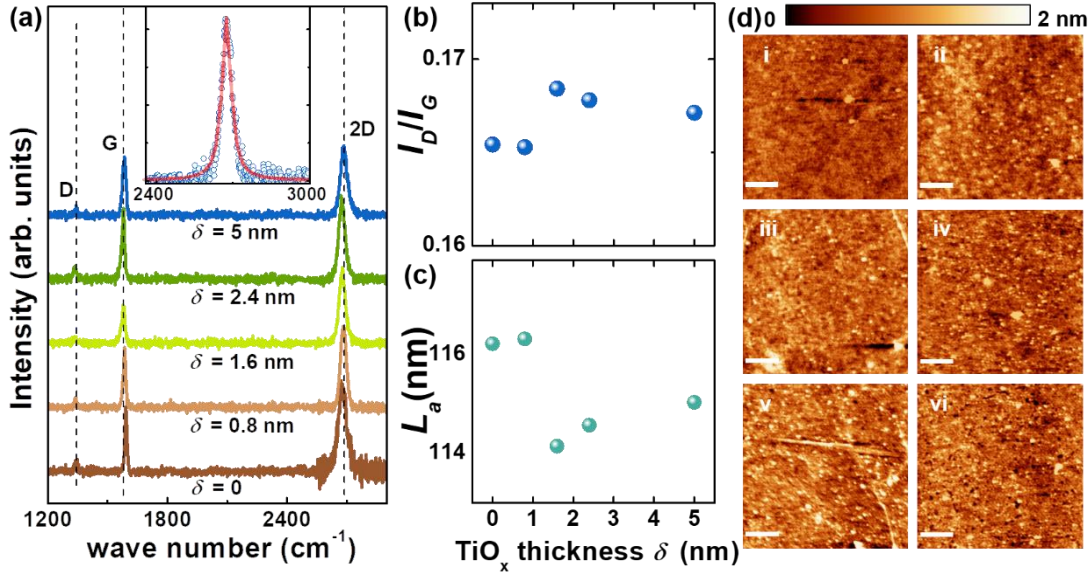

Figure 1: (a) Micro-Raman spectra measured from SLG/TiO$_x$($\delta$)/Co(5 nm) samples (SLG: single-layer graphene). (b) Variation of ratio of $D$ and $G$ peak intensities ($I_D/I_G$) with TiO$_x$ barrier layer thickness $\delta$. (c) Variation of average crystallite size $L_a$ with $\delta$. (d) 2 μm × 2 μm atomic force microscopy (AFM) images of SLG/TiO$_x$($\delta$)/Co(5 nm) samples (i: No SLG, ii: $\delta$=0, iii: $\delta$=0.8 nm, iv: $\delta$=1.6 nm, v: $\delta$=2.4 nm, vi: $\delta$=5 nm). The white scale bar corresponds to a length of 400 nm.

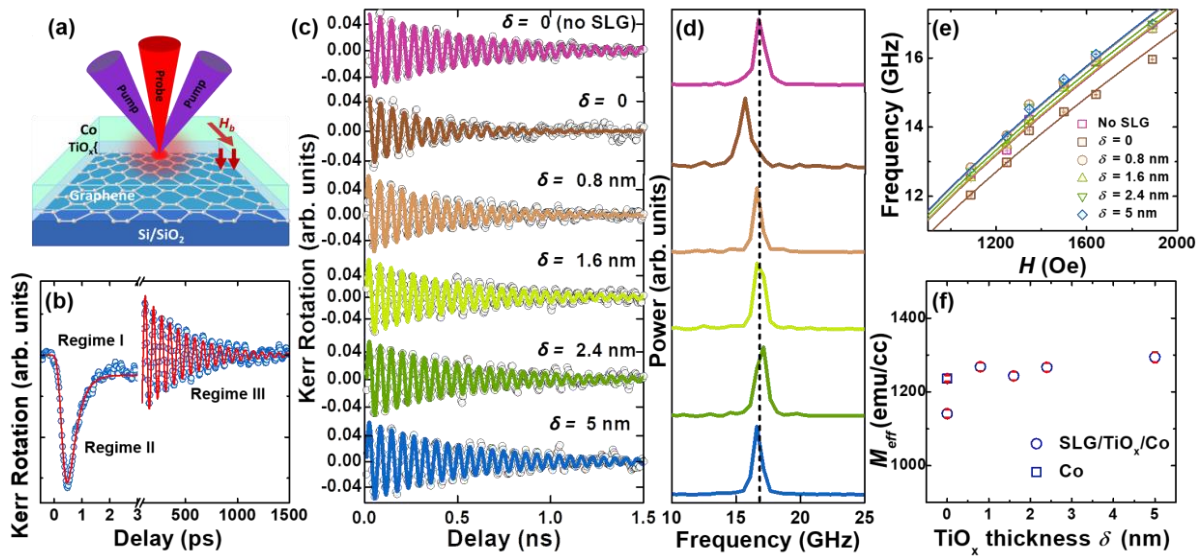

Figure 2: (a) Schematic of two-color pump-probe time-resolved magneto-optical Kerr effect (TRMOKE) measurement geometry. (b) A typical TRMOKE trace depicting: Regime I: ultrafast demagnetization, Regime II: fast magnetization recovery, and Regime III: magnetization precession and damping. (c) Time-resolved Kerr rotation showing magnetization precession and (d) corresponding fast Fourier transform (FFT) power spectra in SLG/TiO$_x$($\delta$)/Co(5 nm) samples. (e) Bias magnetic-field-dependent frequency dispersion for SLG/TiO$_x$($\delta$)/Co(5 nm) samples where solid lines are fits to the Kittel dispersion relation. (f) Dependence of effective saturation magnetization ($M_{eff}$) values on $\delta$.

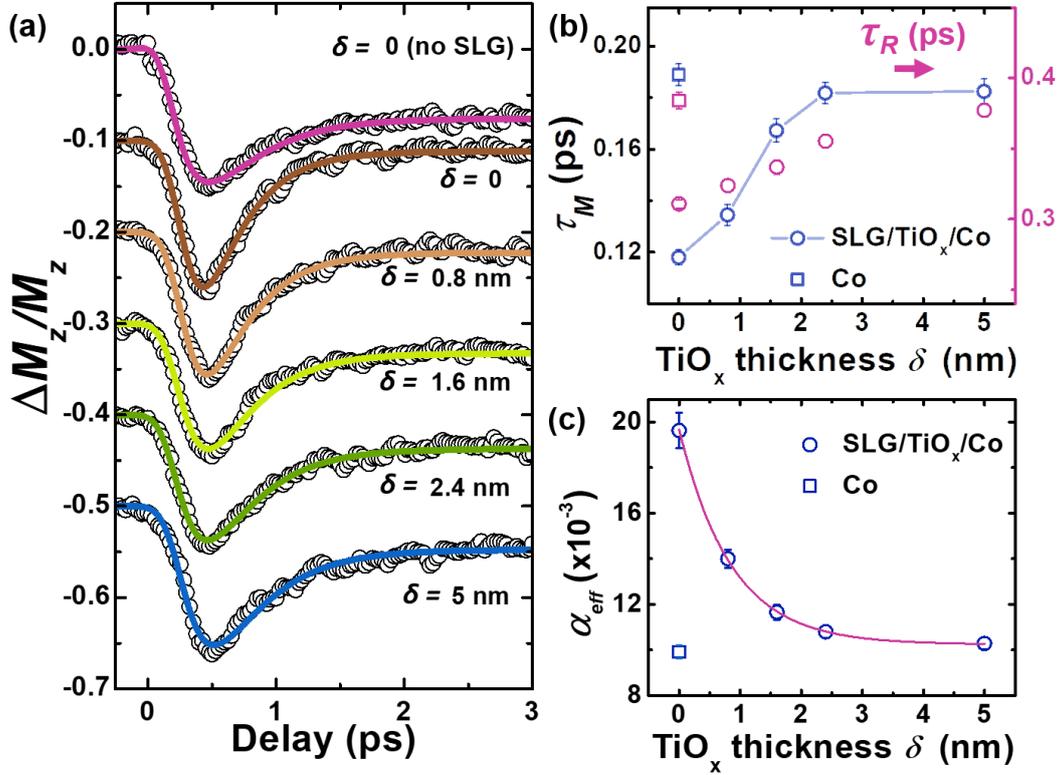

Figure 3: (a) Ultrafast demagnetization traces measured on SLG/TiO$_x$($\delta$)/Co(5 nm) samples. (b) Variation of demagnetization time $\tau_M$ and the fast remagnetization time $\tau_R$ with $\delta$. (c) Variation of Gilbert damping parameter $\alpha_{eff}$ with $\delta$.

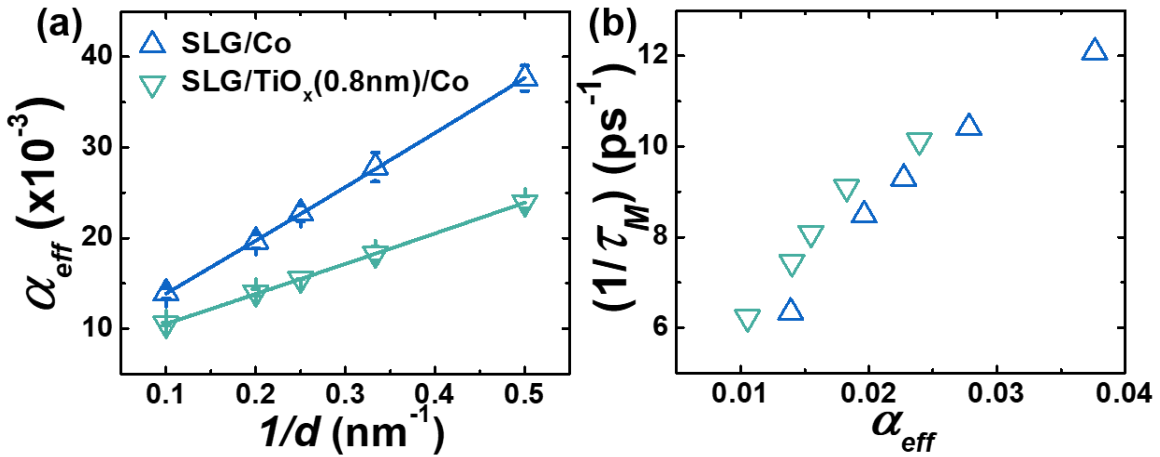

Figure 4: (a) Variation of Gilbert damping parameter $\alpha_{eff}$ with Co thickness in SLG/TiO$_x$($\delta$ = 0, 0.8 nm)/Co($d$) samples where $d$ = 2, 3, 4, 5 & 10 nm. (b) Correlation of ultrafast demagnetization rate with $\alpha_{eff}$.

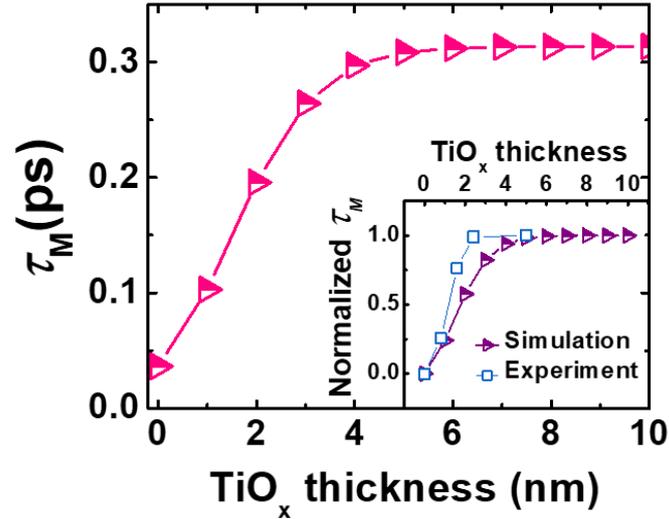

Figure 5: Demagnetization time $\tau_M$ as a function of the TiO$_x$ layer thickness $\delta$, computed with superdiffusive spin-transport theory for Co(5 nm)/TiO$_x$($\delta$). Inset shows normalized variation compared to experimental data.